# Charge-state stability of single NV centers in HPHT-type IIa diamond

Darya Meniailava[1], Michael Petrov[1], Josef Soucek[1,2,3], Milos Nesladek[1,3]

[1]*Institute for Materials Research (IMO), Hasselt University, Wetenschapspark 1, B-3590 Diepenbeek, Belgium.*
[2]*Faculty of Biomedical Engineering, Czech Technical University in Prague, Sítna sq. 3105, 27201 Kladno, Czech Republic.*
[3]*IMOMEC division, IMEC, Wetenschapspark 1, B-3590 Diepenbeek, Belgium.*

Corresponding Author: Darya Meniailava (darya.meniailava@uhasselt.be)

**Abstract**

We investigate the charge-state stability of individual nitrogen–vacancy (NV) centers in weakly doped HPHT IIa diamond containing sub-ppm concentrations of boron and nitrogen. Using Ti/Al coplanar electrodes on an oxygen-terminated surface, we study how applied electric fields and optical excitation jointly govern NV charge conversion. By combining voltage-dependent photoluminescence, real-time charge-state monitoring, laser-power saturation with spectral decomposition, and time-resolved measurements, we reveal that electric fields several micrometers from the contacts significantly increase the $NV^-$ population and enhance spin readout. At low excitation powers, the $NV^-$ population evolves on minute timescales following compressed-exponential kinetics, consistent with slow space-charge rearrangement in ultra-insulating diamond. Under pulsed excitation, we observe microsecond $NV^- \rightarrow NV^0$ conversion driven by hole capture, which is strongly suppressed by applied bias. Our results demonstrate that residual boron acceptors play a key role in determining charge-state stability and show how electrical bias can reliably stabilize $NV^-$ in weakly doped bulk diamond.

**Intoduction**

The nitrogen-vacancy (NV) center in diamond is one of the most widely studied solid-state defects due to its optical addressability, long spin coherence time at room temperature, and prospective applications in quantum communication, nanoscale sensing, and quantum information processing [1,2]. These applications rely on the negatively charged state $NV^-$, whose spin-triplet ground state enables optical initialization, coherent microwave control, and readout through optically detected magnetic resonance (ODMR).

Several approaches have been developed to control the NV charge state by modifying this local electronic environment. Bulk doping with nitrogen, phosphorus, or boron alters the global Fermi-level position and has been used to stabilize selected charge states [3–5]. Surface termination provides another powerful mechanism: oxygen- or fluorine-terminated surfaces can stabilize $NV^-$, whereas hydrogen termination induces a two-dimensional hole layer that promotes conversion to $NV^0$ or $NV^+$ [6,7]. Additional strategies include dielectric or graphene passivation layers, which alter surface band bending and trap densities, and electrical biasing of Schottky or metal-semiconductor structures, where an applied voltage actively shifts the Fermi level and controls the local charge population [8–11]. These approaches illustrate that the NV charge state is highly sensitive to local electrostatics, carrier densities, and band-bending.

However, in ultra-pure or weakly doped diamond, the very low free-carrier concentration strongly limits the applicability of macroscopic band-bending or equilibrium Fermi-level models. In such insulating material, the equilibration of space charge is slow, the density of compensating dopants is low, and long-range electric fields can persist in the diamond bulk. As a result, the charge state of a single NV center may depend not only on surface termination or global doping, but also on the proximity of residual impurities, local potential fluctuations, and slowly evolving space-charge distributions. Understanding how these effects manifest in material with low dopant concentrations is essential for the development of scalable NV-based quantum devices.

In this work, we investigate single NV centers in HPHT IIa diamond with nitrogen and boron concentrations below 300 ppb. We study NV centers located between Ti/Al electrodes fabricated on an oxygen-terminated surface and examine how electrical bias and optical excitation jointly control their charge and spin properties. To this end, we combine steady-state photoluminescence and ODMR measurements under applied voltage, long-time monitoring of NV charge dynamics on minute timescales, laser-power saturation curves with spectral decomposition of $NV^-$ and $NV^0$ contributions, and time-resolved photoluminescence under pulsed excitation. This approach allows us to probe charge-state stability across wide timescale range and to determine how both bias and illumination influence the kinetics of charge conversion in an ultra-insulating diamond environment.

We find that even in diamond with sub-ppm boron and nitrogen content, the electric field generated by the electrodes is sufficient to modify the NV charge state several micrometers from the contacts. Real-time charge-state monitoring reveals a slow, compressed-exponential growth of the NV⁻ population at low excitation powers, indicating that the equilibration of space charge in this ultra-insulating material is cooperative and proceeds on minute-timescales. In contrast, time-resolved measurements show that under pulsed excitation the NV⁻ → NV⁰ conversion occurs on microsecond timescales through hole capture, and that this pathway is significantly suppressed when bias is applied. Taken together, our results show how bias and optical excitation together determine the NV charge state in weakly doped bulk diamond and highlight the important role of residual boron acceptors in determining charge-state stability.

**Results and discussion**

*Voltage-dependent photoluminescence and spin properties*
We investigated HPHT diamond samples from New Diamond Technology containing single NV centers. The concentrations of substitutional nitrogen and boron were < 300 ppb. On the oxygen-terminated diamond surface we fabricated coplanar Ti/Al electrodes, connected to a voltage source and a lock-in amplifier to measure the current (Fig. 1a). A microwave wire, connected to an AWG, was used to control the spin population between $m_s = 0$ and $m_s = \pm 1$.

The confocal map in Fig. 1b shows several resolved NV centers located between the electrodes, which are separated by a 2.5 μm gap. One of these centers, labeled $NV_2$, is positioned approximately in the middle of the gap.

Figure 1c presents the second-order correlation function $g^2(\tau)$ for one of the investigated centers. The value $g^2(0) = 0.22 < 0.5$ confirms that we address single-photon emitters. Figure 1d shows the dependence of the PL count rate on the applied voltage for three NV centers at different positions relative to the electrodes.

The red curve corresponds to NV1, located to the left of the biased electrode (Fig. 1a). For negative voltages, $NV_1$ shows lower PL counts, while for positive voltages the PL increases and reaches a maximum close to zero bias. This trend can be linked to carrier redistribution under the applied electric field, where the sign of the voltage changes the local distribution of electrons and holes. The blue curve shows $NV_3$, which is located on the opposite side, near the electrode connected to the lock-in amplifier. Its behavior is approximately mirrored with respect to the voltage polarity compared to $NV_1$. The black curve represents $NV_2$, situated in the middle of the gap. $NV_2$ shows a symmetric dependence: reduced PL around zero bias and higher PL for both positive and negative voltages.

This pattern is consistent with a metal-semiconductor-metal structure, where each Ti/Al contact forms a Schottky barrier. Around zero bias both barriers are similar, and a relatively high hole density near the electrodes leads to reduced NV⁻ fluorescence. When a sufficiently large positive or negative voltage is applied, one of the barriers is partially suppressed, a depletion region forms, and the local carrier density changes, which modifies the NV charge state. We observed similar voltage-dependent behavior for several NV centers located up to ~20 μm away from the electrodes. The PL response did not follow the voltage instantaneously; instead, it relaxed with a characteristic delay, here limited by the 400 ms step between successive voltage points, indicating slow charge equilibration.

Charge-state dynamics provide a natural explanation for these observations. It is known that charge-state switching between NV⁻ and NV⁰ leads to a reduction in apparent fluorescence intensity [11,12] reported voltage-induced switching between NV⁺, NV⁰ and NV⁻ near electrodes, which appeared as disappearance and reappearance of bright NV spots in confocal images. In our experiment we observe a similar voltage-controlled PL modulation, but for NV centers that are distributed in the bulk rather than shallow centers close to the surface. This suggests that the dominant mechanism is not only related to surface traps but also to the redistribution of bulk carriers.

We attribute the PL suppression near zero bias to the presence of holes from boron acceptors. At low or zero external bias, band bending in diamond can lead to hole accumulation and a Fermi level position that favors $NV^0$ over $NV^-$. Applying a voltage to the electrodes creates a depletion region where the hole density is reduced. This shifts the local Fermi level toward the $NV^-/NV^0$ transition energy, stabilizing the $NV^-$ charge state and increasing the PL count rate.

For $NV_2$, located approximately in the center of the gap, we can use the I-V characteristics to estimate the acceptor concentration. From a characteristic switching voltage of about 3 V and a depletion width of ~1 µm (assuming that $NV_2$ lies in the middle of the depletion region), we obtain an acceptor density on the order of $6 \times 10^{15}$ cm$^{-3}$. This value is consistent with weak p-type doping due to residual boron.

Because its ground state is a spin triplet, the $NV^-$ center shows ODMR. Figure 1e displays CW-ODMR spectra of $NV_2$ for three voltages: -10 V (black), 0 V (red), and +10 V (blue). At 0 V, the ODMR contrast is only about 5%. When we apply ±10 V, the contrast increases to 15-20%, and at the same time the off-resonant PL rises by roughly one third. The solid curves in Fig. 1e are Lorentzian fits with a center frequency around 2870 MHz.

In pulsed ODMR experiments without applied bias we could not detect any spin contrast, while under applied voltage the pulsed results agree with the CW data. The total dark time between initialization and readout did not exceed 0.7 µs, yet at 0 V this was already sufficient for charge dynamics to wash out the difference between on- and off-resonance signals. This indicates that, in the absence of an electric field, the $NV^-$ population is not stable during the dark interval, likely due to fast $NV^- \rightarrow NV^0$ conversion. Under applied bias, the $NV^-$ state is stabilized, and clear spin contrast appears. This behavior is consistent with the fact that only $NV^-$ has a spin-triplet ground state detectable by ODMR, whereas $NV^0$ does not show such resonances.

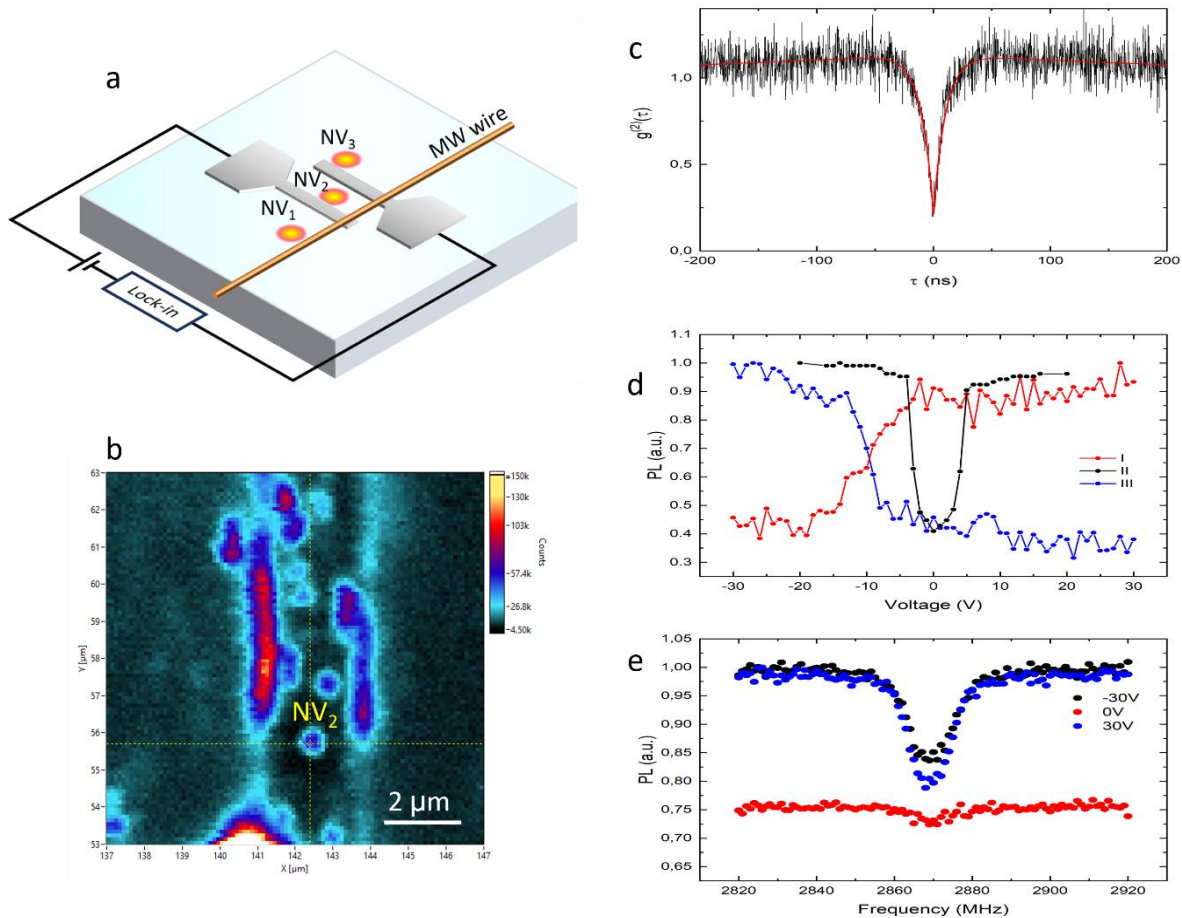

Figure 1. Charge-state control of single NV centers in an HPHT IIa diamond.
*(a)* Sample layout: oxygen-terminated HPHT diamond with coplanar Ti/Al electrodes and a microwave (MW) wire. Electrodes are wired to a voltage source and a lock-in amplifier for simultaneous photocurrent, PL, and ODMR read-out. Colored circles mark the positions of three individually addressed NV centers ($NV_1$ - $NV_3$) relative to the electrodes. *(b)* Confocal PL image of the inter-electrode gap; NV$_2$ center is situated roughly equidistant from both contacts. *(c)* Second-order correlation function g$^2$(τ) for a representative NV center; g$^2$(0) < 0.5 confirms single-photon emission. *(d)* Voltage dependence of continuous-wave PL for the three NVs. NV$_1$ (red, left from the contacts) brightens under positive bias, NV$_3$ (blue, right from the contacts) under negative bias, while centrally located NV$_2$ (black) shows symmetric behavior - this evidence of electrode-controlled charge conversion. *(e)* ODMR spectra of NV$_2$ at 0 V (red), -10 V (black) and +10 V (blue). Bias tuning enhances the spin-read-out contrast by nearly a factor of six, illustrating the link between charge state and spin signal.

*Real-time observation of NV charge conversion under applied bias*

To investigate the charge-state dynamics at low excitation power, we performed real-time photoluminescence (PL) monitoring while applying voltages of 0, 5, 9, and 20 V at the beginning of each measurement. Photon counts were recorded in 100 ms bins, and the total acquisition time varied between 18 and 43 minutes. Representative data are shown in Fig. 2(a) for 9 V and Fig. 2(b) for 0 V, both taken during the fifth minute of the measurement. The lower count level (≈150 cps) corresponds to the NV$^0$ charge state, whereas the higher level (≈270 cps) corresponds to NV$^-$. This clear separation is enabled by the 650 nm long-pass filter, which effectively isolates the NV$^0$ and NV$^-$ emission bands.

The solid red curves in Fig. 2(a,b) show the two most probable bright (NV$^-$) and dark (NV$^0$) states extracted using a Hidden Markov Model. The histograms on the right display the Poisson count distributions accumulated over the full measurement time. Each peak corresponds to a distinct charge state, with numerical labels indicating their relative populations. These histograms reveal that, in the absence of an applied electric field (Fig. 2b), the NV$^-$ state is observed only rarely.

During the monitoring, photon bursts associated with NV$^-$ emission appeared only after some time of laser illumination when voltages of 5-9 V were applied. We analyzed these data minute by minute to extract the temporal evolution of the NV$^-$ and NV$^0$ populations. Figure 2(c) shows the averaged NV$^-$ population as a function of time. The blue curve (0 V) demonstrates that NV$^0$ dominates, while the red, green, and orange curves show the progressive increase of NV$^-$ population with time at 9, 7, and 5 V, respectively. The onset of the first photon burst occurs sooner as the applied voltage increases. At 20 V (violet curve in Fig. 2c), the NV$^-$ population is already nonzero from the first minute of measurement. For each voltage, the experiment was continued until a stable NV$^-$ population of 0.55-0.6 was reached.

The solid red, green, and orange lines in Fig. 2(c) show fits using a compressed exponential function $y = y_0 + A(1 - e^{-(x/\tau)^\beta})$, while for the blue and violet curves, the parameters were fixed using assumptions for τ at 0 V and 20 V to obtain straight reference lines (see Table 1). The characteristic rise time τ decreases strongly with increasing voltage, whereas the shape parameter β remains constant within the experimental error. This behavior indicates that the electric field accelerates the conversion dynamics without altering the underlying kinetic mechanism.

We also verified that both the yellow laser and the applied voltage contribute simultaneously to the observed dynamics. When only one of them was applied (laser or voltage alone), the NV$^-$ bursts appeared only after some certain time, when both sources were turned on together, confirming that optical and electrical excitation are both required to initiate the conversion process.

In our measurements, the two Poisson peaks could not be fully separated, as reported in other works using high numerical-aperture oil-immersion objectives. We deliberately used an air objective to avoid damage to electrodes and diamond surfaces, since oil and cleaning procedures after wire bonding can often degrade the device.

Our diamond is weakly doped (B, N < 300 ppb). Under zero bias we predominantly observe the neutral charge state NV$^0$, suggesting that the Fermi level lies close to the valence band. Applying a positive bias to the Ti-Al electrodes reduces the local hole concentration in the depletion region and raises the Fermi level toward mid-gap, near the NV⁻/NV⁰ transition energy. This shift drives the equilibrium toward the negatively charged NV⁻ state. The minute-scale evolution at low laser power reflects slow space-charge rearrangement in the ultra-insulating diamond, where the near absence of intrinsic carriers makes the charges in the band-bending region effectively immobile [13,14]. As a result, the Fermi level adjusts only slowly.

At higher laser powers, the dynamics become much faster because photogeneration increases the carrier density. Stronger illumination also photo-ionizes acceptors and deep centers, providing additional free carriers that accelerate recombination with NV centers. Therefore, at high power, the NV⁻ population reaches steady state almost instantly [15,16].

The NV⁻ population follows a compressed exponential behavior with $\beta \approx 5\text{-}7 > 1$ for all bias voltages. Such compressed exponential growth suggests cooperative kinetics rather than independent random events. Once a critical level of hole depletion is achieved, the local potential around the NV center changes rapidly, stabilizing NV⁻ in an avalanche-like process. These sigmoidal kinetics resemble Avrami-type models used for cooperative phase transformations [17] and field-induced switching in ferroelectrics [18]. In our case, the cooperative mechanism likely arises from collective space-charge modification within the depletion region: each charged defect or NV⁻ partially screens the local field, facilitating further NV charging once the process begins. The independence of $\beta$ on voltage supports this interpretation, indicating that the mechanism remains the same, while the applied electric field primarily determines the timescale $\tau$ (see Table 1).

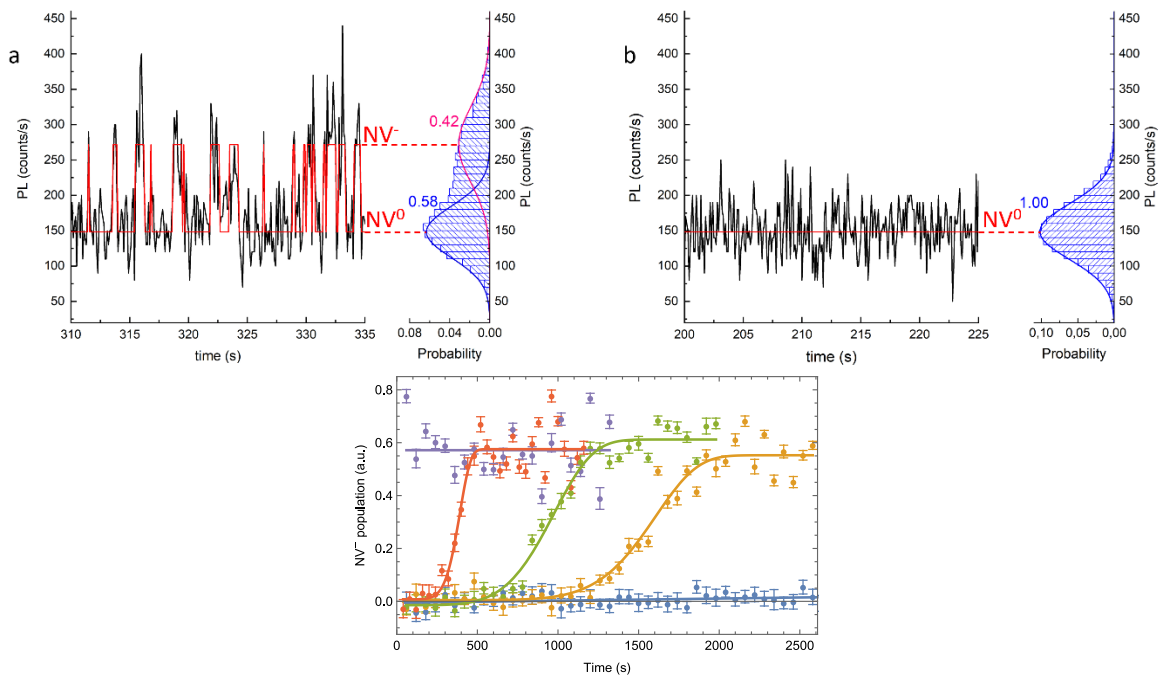

Figure 2. Real-time monitoring of voltage-controlled charge conversion in a single NV center. *(a, b)* PL traces at +9 V (a) and 0 V (b). High plateaus correspond to NV⁻, low plateaus to NV⁰. Red line is a two-state Hidden Markov-model fit. Count-rate histograms show Poisson peaks for NV⁻ (bright) and NV0 (dim), yielding their relative populations. *(c, d)* Dwell-time of the NV⁻ state (c) and NV⁰ state (d) under +20 V (green), +9 V (black), +7 V (red), and +5 V (blue) bias as function of time. Increasing bias shortens NV⁰ lifetimes while prolonging NV⁻ lifetimes, confirming field-assisted charge stabilization.

Table I. Parameters of compressed exponential fits describing NV⁻ population dynamics at different bias voltages.

| Bias (V) | y₀ | A | τ (s) | β |
|---|---|---|---|---|
| 0 | 0.001 ± 0.004 | | ∞ | |
| 5 | 0.00 ± 0.01 | 0.55 ± 0.02 | 1639 ± 29 | 7.1 ± 1.1 |
| 7 | 0.00 ± 0.02 | 0.61 ± 0.02 | 1022 ± 20 | 5.4 ± 0.7 |
| 9 | 0.00 ± 0.03 | 0.57 ± 0.04 | 401 ± 14 | 6.9 ± 2.2 |
| 20 | | 0.57 ± 0.02 | 0 | |

*Laser-Power-Dependent Charge Dynamics and Spectral Decomposition*

The dependence of the photoluminescence (PL) count rate on the applied laser power is shown in Fig. 3(a). The red curve corresponds to $NV_2$ under a 10 V bias applied to the electrodes, whereas the black curve shows the response of the same NV center at 0 V. The solid lines represent steady-state fits obtained from the rate-equation model. Under a 10 V bias, the PL exhibits a clear maximum at approximately 4 mW, followed by a gradual decrease at higher powers. In contrast, the 0 V curve follows a typical saturation behavior, steadily approaching its maximum value. The two curves cross at 5.6 mW: below this power, the count rate is lower at 0 V than at 10 V, consistent with the behavior observed in Fig. 1(d) (black curve). At high laser power, the NV center is approximately one-third brighter without the applied electric field. All curves include background correction, accounting for the linear increase of background counts with excitation power.

To determine the relative contributions of $NV^0$ and $NV^-$ to the saturation behavior, we recorded emission spectra at several laser powers. Figure 3(b) shows representative spectra for 0 V (black) and 10 V (red). The spectrum at 10 V contains a narrow Raman line at 572 nm and clear signatures of $NV^-$ emission: the zero-phonon line (ZPL) at 637 nm and its broad phonon sideband extending from 650 to 750 nm. In addition to these features, the 0 V spectrum displays the $NV^0$ ZPL at 575 nm and its sideband between 580 and 637 nm, indicative of a significant $NV^0$ contribution. Beyond 637 nm, the $NV^0$ sideband overlaps with the $NV^-$ ZPL and $NV^-$ sideband, making the charge-state separation non-trivial without spectral decomposition.

The results of spectral decomposition are presented in Fig. 3(c). The $NV^-$ reference spectrum was taken from the 10 V dataset at 1.1 mW [Fig. 3(b), red curve]. The $NV^0$ reference was obtained by a weighted subtraction of the 0 V and 10 V spectra in Fig. 3(b). The data points in Fig. 3(c) represent the integrated spectral area of either $NV^0$ or $NV^-$ emission within 550–750 nm, normalized to the maximum $NV^-$ area (recorded at 0 V and 43 mW). Filled symbols denote the $NV^-$ contribution, and open symbols denote $NV^0$; black and red colors correspond to 0 V and 10 V, respectively.

At low excitation powers, the $NV^0$ contribution at 0 V is unexpectedly high—for example, at 1.1 mW the $NV^0$ fraction reaches 0.6. This finding is consistent with the reduced PL counts in the saturation curve [Fig. 3(a), black], the lower off-resonant fluorescence in the ODMR measurement [Fig. 1(e), red], the reduced PL in the voltage-dependence traces [Fig. 1(d)] for NV1, NV2, and NV3, and the small spin contrast of only 5% observed for NV2 [Fig. 1(e), red].

Under a 10 V electric field, the $NV^0$ contribution (red open circles) remains low up to ~12 mW and then gradually increases with power. In contrast, the 0 V data exhibit a more rapid decrease of $NV^0$ fraction, reaching a minimum between 6 and 12 mW, followed by a re-emergence of $NV^0$ at higher powers. Notably, the $NV^0$ contribution is larger at 0 V for both low and high laser powers; only around ~6 mW—where the saturation curves intersect [Fig. 3(a)]—do the $NV^0$ fractions coincide. The $NV^-$ areas (filled symbols) closely follow the trends of the PL saturation curves measured with the 650 nm long-pass filter. At high excitation powers, the $NV^-$ emission is significantly brighter in the absence of an electric field.

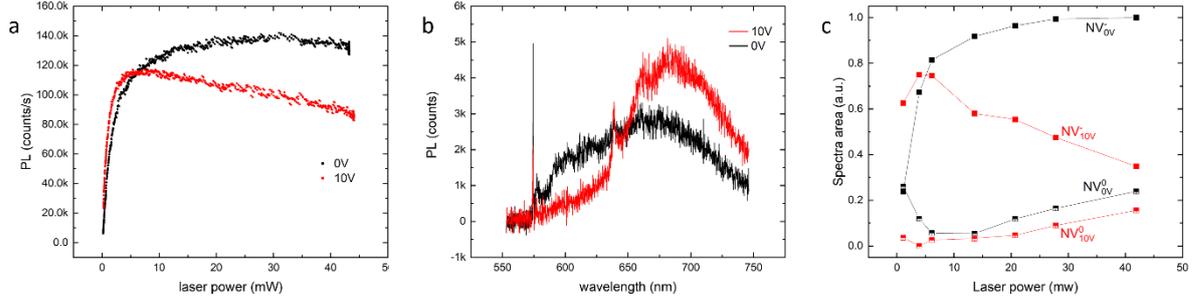

Figure 3. Bias-dependent saturation behavior and spectral fingerprint of NV charge conversion. *(a)* Laser-power saturation curve of a single NV center at 0 V (black) and +10 V (red). Higher count rate evidences a larger $NV^-$ population. *(b)* Room-temperature photoluminescence spectra ($\lambda_{exc}$ = 532 nm, 1 mW) recorded under the same two bias conditions. The +10 V trace shows a pronounced $NV^-$ zero-phonon line at 637 nm and a suppressed $NV^0$ line at 575 nm. *(c)* Integrated areas obtained from spectral deconvolution: filled squares represent the $NV^-$ contribution, half-filled squares the $NV^0$ contribution. The graph highlights the bias-driven shift from a mixed $NV^-/NV^0$ state at 0 V to a predominantly $NV^-$ state at +10 V, corroborating the optical saturation data.

*Time-resolved PL dynamics*

Figure 4(a) shows time-resolved photoluminescence (PL) from a single NV center excited by a 2 μs green laser pulse at 1.1 mW. The black curve was recorded with no voltage applied, and the red one with +10 V between the electrodes. When the laser is switched on, the red trace shows a short and strong PL spike that decays to an almost constant level within the 2 μs pulse, while the black trace shows a slower rise of PL intensity until it reaches equilibrium. The steady-state count rate is higher under +10 V, which agrees with the bias dependence of the $NV^-$ population shown in Fig. 1(d-e) and Fig. 3(a). At the end of the laser pulse, the counts drop immediately, with a fast decay time of about 10 ns. After the laser is turned off, a weak residual PL remains visible and decays on a microsecond timescale. At 0 V, the decay time is about 457 ± 15 ns. The inset in Fig. 4(a) compares these dark decays for 0 V and +10 V together with exponential fits.

The different shapes of the PL curves in Fig. 4(a) reflect two distinct regimes of NV dynamics. The pronounced spike in the NV emission (red curve) immediately after the green laser is switched on, followed by an exponential decay with a characteristic time constant of 195 ± 22 ns at low laser powers, indicates that the NV center initially resides in the negatively charged $NV^-$ state, which emits strongly upon optical excitation. The subsequent decay reflects the internal relaxation dynamics of the $NV^-$ center rather than charge-state conversion. In such dynamics the transient decrease of PL intensity is explained by spin-dependent intersystem crossing: after excitation to the triplet state, part of the population is transferred non-radiatively into the long-lived singlet manifold, which relaxes back to the triplet ground state on a timescale of a two-three hundred nanoseconds [1], [19], [20]. This sequence of transitions naturally produces a short PL spike at laser turn-on followed by a decay toward steady state, as we observed.

When the laser pulse is applied at 0 V (black curve), the PL intensity gradually increases with a time constant of $\tau$ = 270 ± 7 ns (at 1 mW), extracted from a double-exponential fit (second exponent represents already mentioned spike decay of NV-). We should note that this time constant decreases with increasing laser power. Such rise of the signal indicates that the NV center is initially almost in the neutral $NV^0$ state and gradually transforms into the negatively charged $NV^-$ state under green excitation. The use of a 650 nm long-pass filter ensures that the detected signal originates predominantly from $NV^-$ emission. The observed optical charging process is well known from earlier studies of NV charge dynamics, where green illumination induces photo-ionization and electron capture processes between $NV^0$, $NV^-$, and nearby traps or donors [21]. A similar charging rise on sub-microsecond to microsecond scales depending on laser power was also observed and attributed to sequential photo-ionization of $NV^0$ followed by electron capture from the valence band or from nearby donors [22].

To further understand the relaxation after the laser pulse, we analyzed how the PL decay time in the dark depends on laser power. Figure 4(b) summarizes these results. Each point shows the decay time τ obtained from exponential fits of the post-pulse PL, and the solid line represents a hyperbolic fit. The decay time becomes shorter with increasing laser power and approaches a constant value of 272 ± ns. This behavior indicates that the decay is governed by hole capture by $NV^-$ centers: higher optical power generates more holes, which are captured faster, leading to τ ∝ 1/P. The constant term $\tau_0 \approx 272$ ns represents the fastest possible decay when hole capture is no longer the limiting step.

The offset value $\tau_0$ can be explained by the lifetime of the metastable singlet state in the $NV^-$ optical cycle, which is of a few hundred nanoseconds at room temperature [19], and our own measurements give 195 ± 22 ns (from the spike fit) and 210 ± 22 ns, as measured using the same protocol in [19]. The value is 20-30% smaller than fitted 272 ns but close. This suggests that, at high power, the PL tail is limited mainly by relaxation through the singlet state. Another possible reason for the offset could be partial saturation of photo-generated holes at high power, but such an effect would normally distort the simple 1/P trend, while our data follow it very well. We therefore attribute the offset mostly to the intrinsic $NV^-$ singlet lifetime, with hole saturation playing only a minor role.

This interpretation agrees with other studies of carrier-assisted charge conversion. In [23] very large hole-capture cross-sections for $NV^-$ was found, confirming that NV- acts as an efficient hole trap. In a related work [24] it was shown that photo-generated carriers created on nearby defects can diffuse and be captured by a probe $NV^-$, changing its fluorescence. In [25] it was also demonstrated that NV charge-cycling strongly depends on the rate of carrier generation. Our data are therefore consistent with this general picture of carrier-limited NV charge dynamics, where higher illumination produces more holes that are captured faster by $NV^-$, resulting in shorter decay times.

To determine which charge state emits in the dark, we studied the spectral dependence of the PL tail. Time-resolved signals were recorded in three spectral windows: 550-700, 600-700, and 650-700 nm. We integrated the delayed PL between 4 μs and 14 μs after the laser pulse and normalized it to the maximum value at 550-700 nm. The same procedure was applied to the steady-state $NV^-$ and $NV^0$ spectra integrated over the same ranges. Figure 4(c) compares the normalized tail area (x-axis) with the normalized spectral area (y-axis) for $NV^-$ (red) and $NV^0$ (black). The dashed line shows the ideal correlation (y = x).

The $NV^0$ data lie close to the y = x line, meaning that the delayed PL scales with the $NV^0$ spectral weight, while $NV^-$ shows a much weaker correlation. This proves that the residual PL in the dark mainly originates from $NV^0$ emission.

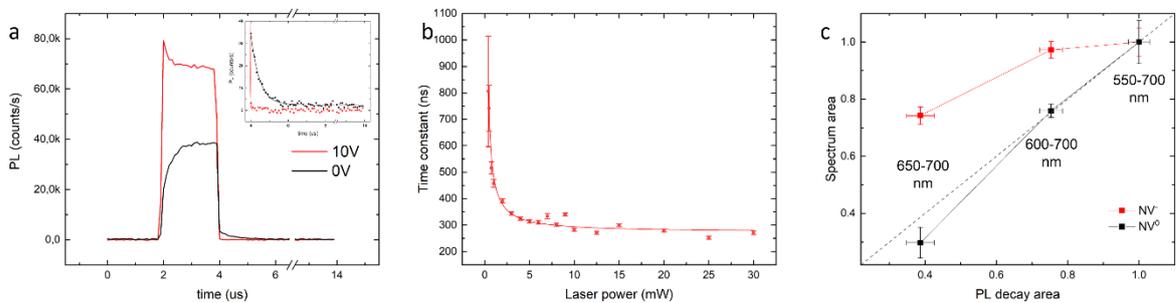

Figure 4. Time-resolved evidence for a dark $NV^- \rightarrow NV0$ transition. (a) Time-resolved PL from $NV_2$ under a 2 μs laser pulse (λ = 532 nm, 1 mW) at 0 V (black) and +10 V (red). Applied bias voltage increases the initial $NV^-$ signal and eliminates a weak tail that persists after the laser is switched off. Inset: magnified 4-14 μs window with single-exponential fits (solid lines) used to extract the tail decay constant τ. (b) τ versus excitation power. The hyperbolic fit (solid line) follows τ ∝ 1/P, indicating that the decay is governed by hole capture whose rate rises with photo-generated carrier density. (c) Correlation between the integrated tail PL (x-axis) and integrated spectrum areas (y-axis) of the NV⁻ (red) and NV0 (black) in three different spectral windows. The tail signal scales with the $NV^0$ component but not with NV⁻, confirming that the long decay originates from the dark NV⁻ → $NV^0$ conversion process.

**Methods**

We investigated HPHT-grown diamond samples from NDT containing single NV centers (Fig. 1a). The concentrations of substitutional nitrogen and boron were < 300 ppb. Prior to electrode fabrication, the sample surface was cleaned in a boiling $H_2SO_4$ + $KNO_3$ solution to ensure a well-defined oxygen termination. Coplanar Ti/Al electrodes were then deposited on the oxygen-terminated surface and connected to a voltage source and a lock-in amplifier for current readout. A microwave wire placed near the NV sites was driven by an arbitrary waveform generator (AWG) to control population transfer between the $m_s = 0$ and $m_s = \pm 1$ spin states. All experiments were carried out using a home-built confocal microscope.

The excitation laser beam was guided to the microscope objective using a series of mirrors and lenses. The sample was mounted on a three-axis piezoelectric stage, enabling precise nanometer-scale positioning. Focusing was achieved with a high-numerical-aperture air objective. Photoluminescence (PL) from the NV centers was collected through the same objective, spectrally filtered by a dichroic mirror and a 650 nm long-pass filter, and detected using a single-photon counting module (SPCM). The AWG also triggered the acousto-optic modulator (AOM, 30 ns rise time) to generate laser pulses, while a National Instruments data-acquisition card was used to read out photon counts.

Second-order autocorrelation measurements $g^{(2)}(\tau)$ were performed using a standard Hanbury Brown–Twiss interferometer consisting of a 50:50 beamsplitter, two SPCMs, and a time-correlation module. Unless otherwise specified, all experiments were conducted with 532 nm green excitation at 1 mW (measured before the objective) and using a 650 nm long-pass filter.

PL spectra were recorded under 532 nm excitation at various laser powers using a 550 nm long-pass filter. The emission was redirected from the detector path to a spectrometer via a movable mirror. Time-resolved measurements of NV luminescence in darkness were performed using a 700 nm short-pass filter combined with long-pass filters with 550, 600, and 650 nm cut-on wavelengths.

Real-time PL monitoring experiments were carried out under 561 nm excitation at 1 μW using a 650 nm long-pass filter, following the procedure described in [15].